\documentclass[fleqn,usenatbib]{mnras}

\usepackage{newtxtext,newtxmath}

\usepackage[T1]{fontenc}
\usepackage{ae,aecompl}
\usepackage{datatool}

\usepackage{graphicx}	
\usepackage{amsmath}	
\newcommand\textlcsc[1]{\textsc{\MakeLowercase{#1}}}

\title[Dark matter and the Galactic Centre Excess]{Dark matter annihilation and the Galactic Centre Excess}

\author[R. J. J. Grand et al.]{\parbox[t]{\textwidth}{
Robert J. J. Grand$^{1,2}$\thanks{E-mail: rgrand@iac.es}, Simon D. M. White$^3$}
\\ 
$^1$Instituto de Astrof\'isica de Canarias, Calle Vía L\'actea s/n, E-38205 La Laguna, Tenerife, Spain\\
$^2$Departamento de Astrof\'isica, Universidad de La Laguna, Av. del Astrof\'isico Francisco S\'anchez s/n, E-38206, La Laguna, Tenerife, Spain\\
$^3$Max-Planck-Institut f\"{u}r Astrophysik, Karl-Schwarzschild-Str. 1, 85748 Garching, Germany\\
}

\date{Accepted XXX. Received YYY; in original form ZZZ}

\pubyear{2022}

\begin{document}
\label{firstpage}
\pagerange{\pageref{firstpage}--\pageref{lastpage}}
\maketitle

\begin{abstract}
We compare the surface brightness profile and morphology of the Galactic Centre Excess (GCE) identified in wide-angle $\gamma$-ray maps from the Fermi-Large Area Telescope to dark matter annihilation predictions derived from high-resolution $\Lambda$CDM magnetohydrodynamic simulations of galaxy formation. These simulations produce isolated, disc-dominated galaxies with structure, stellar populations, gas content, and stellar and halo masses comparable to those of the Milky Way. For a specific choice of annihilation cross-section, they agree well with the Fermi-LAT data over the full observed angular range, $1^{\circ}$ to $15^{\circ}$, whereas their dark-matter only counterparts, lacking any compression of the inner halo by the gravitational effects of the baryons, fail to predict emission as centrally concentrated as observed. These results provide additional support to the hypothesis that the GCE is produced by annihilating dark matter. If, however, it is produced by a different mechanism, they imply a strong upper limit on annihilation rates which can be translated into upper limits on the expected $\gamma$-ray flux not only from the inner Galaxy but also from any substructure, with or without stars, in the Galactic halo.
\end{abstract}




\section{Introduction}

Dark matter (DM) accounts for more than $80\%$ of all matter in the Universe, but its nature is unknown. Historically, a particularly well motivated candidate for the DM has been a Weakly Interacting Massive Particle, perhaps the lightest supersymmetric partner of the known particles \citep[a WIMP, e.g.][]{BHS05}. Such particles may produce observable electromagnetic radiation through annihilation, which, for standard WIMPS, produces $\gamma$-ray emission at GeV energies \citep[see e.g.][]{Arcadi+18}. The indirect detection of DM through this channel has been the subject of much research, but no clear signal, unambiguously due to DM, has so far been confirmed \citep{G16}.

The Large Area Telescope (LAT) on board the Fermi Gamma-ray Observatory provides detailed observations of $\gamma$-ray emission across the entire sky \citep{Fermi09} and has been used to study possible DM annihilation signals from dwarf satellites of the Milky Way \citep[e.g.][]{Albert+17} and from the diffuse component of its main DM halo \citep[e.g.][]{BBM94,BUB98,Ackermann+12,CLM18}. Previous work had argued that the former are the most promising targets to detect an annihilation signal \citep[e.g.][]{Strigari2007}. However, numerical simulations demonstrate that the signal from the Milky Way should be overwhelmingly dominated by smooth background emission, even when the entire subhalo mass function (down to Earth mass haloes) is taken into account \citep{SWF08,WBF20,GW21}. A diffuse emission component in the Galactic Centre region with morphology, radial profile and spectrum similar to the expectations for annihilating DM was first isolated in the LAT data more than a decade ago \citep[][see also arXiv:0910.2998]{Hooper2011}. While it is clear that this Galactic Centre Excess (GCE) is more extended perpendicular to the Galactic Plane than expected for stellar or gaseous disk sources, it might be associated some kind of bulge stellar population. Large uncertainties in the spatial distribution of contamination from other sources of emission make it difficult to characterise the GCE precisely enough to separate the bulge and DM annihilation interpretations.  Pixel-to-pixel variations in the surface brightness of the GCE have been argued to be inconsistent with a DM origin, favouring emission  by a previously unknown population of point-sources in the bulge, for example, faint millisecond pulsars \citep{Lee2016}, but more recent work has demonstrated that these variations are likely to reflect imperfections in the foreground templates rather than a property of the GCE itself; thus the interpretation of the GCE remains uncertain \citep[see][chapter 6 for a thorough recent review of these topics]{Slatyer2021}.

Recently, \citet{DiMauro21} characterised the GCE using a full 11 years of Fermi-LAT data, together with a variety of updated source catalogues and interstellar emission models.  In particular, he presented a map of the GCE surface brightness over a $40^{\circ}\times 40^{\circ}$ region surrounding the Galactic Centre and binned it up to obtain a circularly averaged profile which was consistent with the simplest DM predictions; he estimated the axial ratio of the GCE emission, finding ``best" values in the range 0.8 to 1.2, depending on foreground model, hence consistent with a spherically symmetric or slightly flattened halo; and he showed the GCE energy spectrum to be consistent with several DM annihilation models. He did not, however, revisit the photon statistics issue to check for consistency with the smooth emission expected for DM annihilation radiation. In this paper, we compare Di Mauro's results to predictions from  high-resolution, magnetohydrodynamical simulations of galaxy formation that we previously used to study baryonic effects on the annihilation radiation of Milky Way-like galaxies and their substructures \citep{GW21}. In general, we find remarkably good agreement, supporting the idea that the GCE is dominated by annihilation radiation, possibly from a WIMP, within the $\Lambda$CDM paradigm.

\section{Simulations}
\label{sim}

\begin{figure*}
\includegraphics[width=\columnwidth,trim={0cm 0.2cm 0.5cm 0.5cm}, clip]{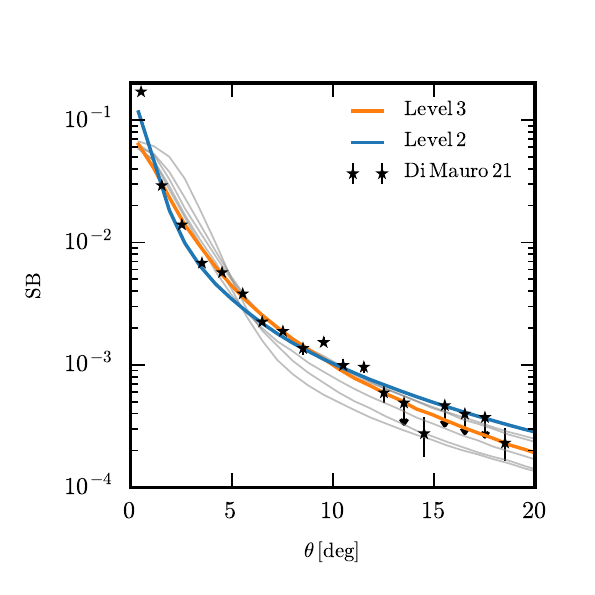}
\includegraphics[width=\columnwidth,trim={0cm 0.2cm 0.5cm 0.5cm}, clip]{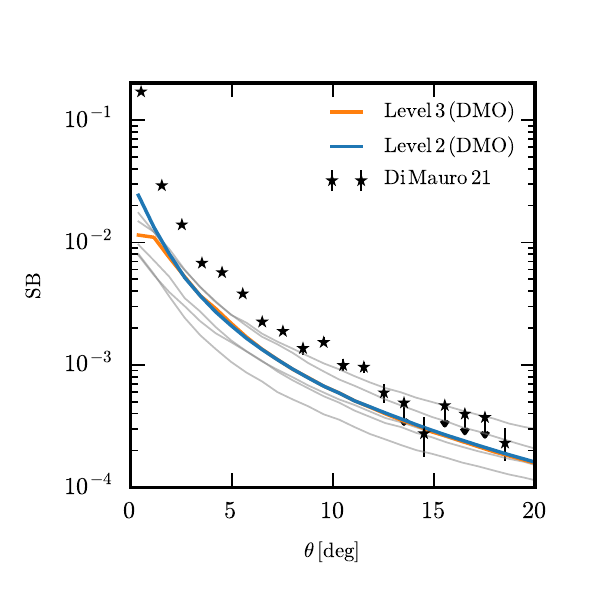}
\caption{Mean angular surface brightness profiles of the emission produced by dark matter annihilation for our simulated haloes (solid curves) in the full physics runs (left panel) and in the dark-matter-only runs (right panel), together with the observed profile of the Galactic Centre Excess (star symbols, note that four of these are upper limits) from the 11 year Fermi-LAT dataset \citep{DiMauro21}. The profile of our highest resolution (level 2) simulation is coloured blue  and its lower resolution (level 3) counterpart is coloured orange; the other five level 3 simulations are coloured  grey. }
\label{ASB}
\end{figure*}

We analyse the same suite of six simulations of Milky Way-like systems and their local environments as in \cite{GW21}.  Taken from the \textlcsc{Auriga} project \citep{GGM17,GHF18}, these systems were selected to have halo masses between $1$ and $2\times 10^{12} \rm M_{\odot}$,\footnote{We define halo mass, $M_{200}$, as the mass within the radius $R_{200}$ that encloses a mean density 200 times the critical value for closure} and to be moderately isolated; they were identified in the $z=0$ snapshot of a DM-only simulation of a periodic cube of comoving size 100 Mpc, assuming a standard $\Lambda$CDM cosmology. The adopted cosmological parameters were $\Omega _m = 0.307$, $\Omega _b = 0.048$, $\Omega _{\Lambda} = 0.693$ and a Hubble constant of $H_0 = 100 h$ km s$^{-1}$ Mpc$^{-1}$, where $h = 0.678$, taken from \citet{PC13}. At $z=127$, the DM resolution  of each halo and its surroundings is increased, and gas is added to create the initial conditions for a new ``zoom" simulation; this is evolved to the present day using the magnetohydrodynamics code \textlcsc{AREPO} \citep{Sp10}. Each simulation is available in a ``full physics" (FP) and in a ``DM-only" (DMO) version.

Galaxy formation processes included in the FP versions of these simulations include self-gravity of all components, dissipative hydrodynamics, radiative cooling of gas, a two-phase model for cold and hot gas in star-forming regions, star formation, mass and metal return from stellar evolution, supermassive black hole formation, accretion and merging, energetic feedback from stars and AGN, and magnetic fields \citep[see][for a full description]{GGM17}. The \textlcsc{AURIGA} model produces disc-dominated, star-forming spiral galaxies that are broadly consistent with a number of observations, in particular, star formation histories, stellar masses, sizes and rotation curves of nearby Milky Way-like galaxies \citep{GGM17}. More detailed properties of such galaxies are also matched, including some particularly relevant to the Milky Way, for example: the distribution of HI gas \citep{MGP16}; the size and structure of their bulges \citep{GMG19,Fragkoudi+Grand+Pakmor+19}; and a thick and thin disc with chemical and structural properties similar to that of the Milky Way \citep{GBG18,GKB20}. These simulations are thus good ``full physics" versions of a $\Lambda$CDM Milky Way, and are hence well-suited for our purposes. 

In the FP runs these systems have mass resolution
$\sim 6 \times 10^3$ $\rm M_{\odot}$ and $\sim 5 \times 10^4$ $\rm M_{\odot}$ for baryons and DM, respectively, while in the DMO runs the DM particle mass is $\sim 6 \times 10^4$ $\rm M_{\odot}$. In both cases, the gravitational softening length is $184$ pc after $z=1$ and is fixed in comoving units at earlier times. For consistency with earlier work, these resolution specifications are referred to as ``level 3''. In addition to these six level 3 simulations, we include here the recently completed ``level 2'' simulation of \citet{GMP21} (both the FP and the DMO versions), for which which the resolution is increased by factors of eight in mass and two in spatial scale.

\section{Results}
\label{results}

To predict annihilation surface brightness distributions from our simulations, we first calculate the luminosity density $\mathcal{L}$ at each point in space using 
\begin{equation}
\mathcal{L}=\mathcal{C}\rho ^2,
\label{lumden}
\end{equation}
where $\mathcal{C}$ depends on observational and particle physics quantities, such as the observing band-pass, the available annihilation channels and their coupling to photon emission, and the velocity-weighted annihilation cross-section, $\langle\sigma {\rm v}\rangle$, We assume these to be position-independent and that the latter is velocity-independent (corresponding to s-wave annihilation).  Thus, up to a multiplicative constant, our calculations boil down to evaluating the DM density at every point in space. Following \citet{GW21}, we apply \textlcsc{AREPO}'s Voronoi tesselator to the DM particle distribution, allowing calculation of $\rho _i$, the density in the cell surrounding the $i$-th DM particle, from the cell mass and volume. This approach provides a better localised measure of the dark matter density than kernel-weighted estimates that smooth over a particle's neighbours. This is especially important for high density regions such as the Galactic Centre.

\subsection{1D angular surface brightness profiles}

With the above assumptions, the annihilation luminosity associated with the $i$-th DM particle is simply the product of its mass $m_i$ and the density $\rho _i$ of its Voronoi cell. To construct annihilation surface brightness (ASB) profiles, we adopt 18 ``Solar" positions equally spaced around an 8 kpc circle in the disk plane of each simulated galaxy, and for each we calculate the projected annihilation flux for a grid of sky pixels within $20^{\circ}$ of the Galactic centre: $f_j = \Sigma _i ^{N_j} \rho _i m_i / d_i^2$, where $N_j$ is the number of DM particles projected onto the $j$-th pixel. We then construct ASB profiles for each position by averaging the surface brightness in annuli around the Galactic Centre. Fig.~\ref{ASB} shows these profiles both for our FP simulations (left) and for our DMO simulations (right) and compares them with the recent ASB profile obtained from 11 years of Fermi-LAT data by \citet{DiMauro21}. For each simulation we have averaged the profiles from the 18 different ``solar" positions, although the scatter among these is typically substantially smaller than the variation between simulations. A normalisation factor has been applied to the simulated ASB curves in order to facilitate comparison of their {\it shapes} to the observed profile; the value of the normalisation (equivalent to the constant, $\mathcal{C}$ in equation \ref{lumden}) is kept the same for all curves in both panels of Fig. \ref{ASB}. 

Focusing first on the full physics predictions, the grey and orange curves in the left panel of Fig.~\ref{ASB} show mean ASB profiles for our six level 3 simulations. The shapes and scatter of these curves are consistent with the observed profile for all but the innermost $\sim 2$ degrees, where they are systematically low. The size of this region corresponds approximately to the gravitational softening of these simulations, so that densities are underestimated within this radius. To illustrate this, we also plot the single
level 2 simulation introduced in \citet{GMP21}, for which the resolution is improved by factors of 8 and 2 in mass and spatial scale, respectively (the blue curve). The level 3 counterpart of this simulation is shown in orange. The mean ASB profile of this high-resolution simulation agrees remarkably well with the observations, noticeably better than its level 3 counterpart at the innermost point. 

The right panel of Fig.~\ref{ASB} shows mean ASB profiles for the DMO counterparts of the FP simulations of the left panel. These are shallower than the corresponding FP profiles and do not fit the observed shape as well; the increase from the outermost measured points to the innermost point is a factor of $\sim 10$ below that observed, even for the high-resolution simulation, which is enhanced by a factor of 2 at the centre relative to its lower resolution counterpart but otherwise agrees with it remarkably well, much better than in the FP case. This difference, like the fact that the FP curves lie systematically above their DM counterparts, and by larger factors in the inner regions, is a consequence of the gravitational compression of the halo during the formation of the visible galaxy and of the substantial variability of this process between realisations and resolution levels \citep[see also][]{SFB15,LPG18,CBD20}. Baryonic contraction of the dark halo appears necessary to reproduce the observed ASB profile of the GCE.

\subsection{The shape of surface brightness isophotes}

\begin{figure}
\includegraphics[scale=0.6,trim={2.5cm 2cm 2.cm 2cm}, clip]{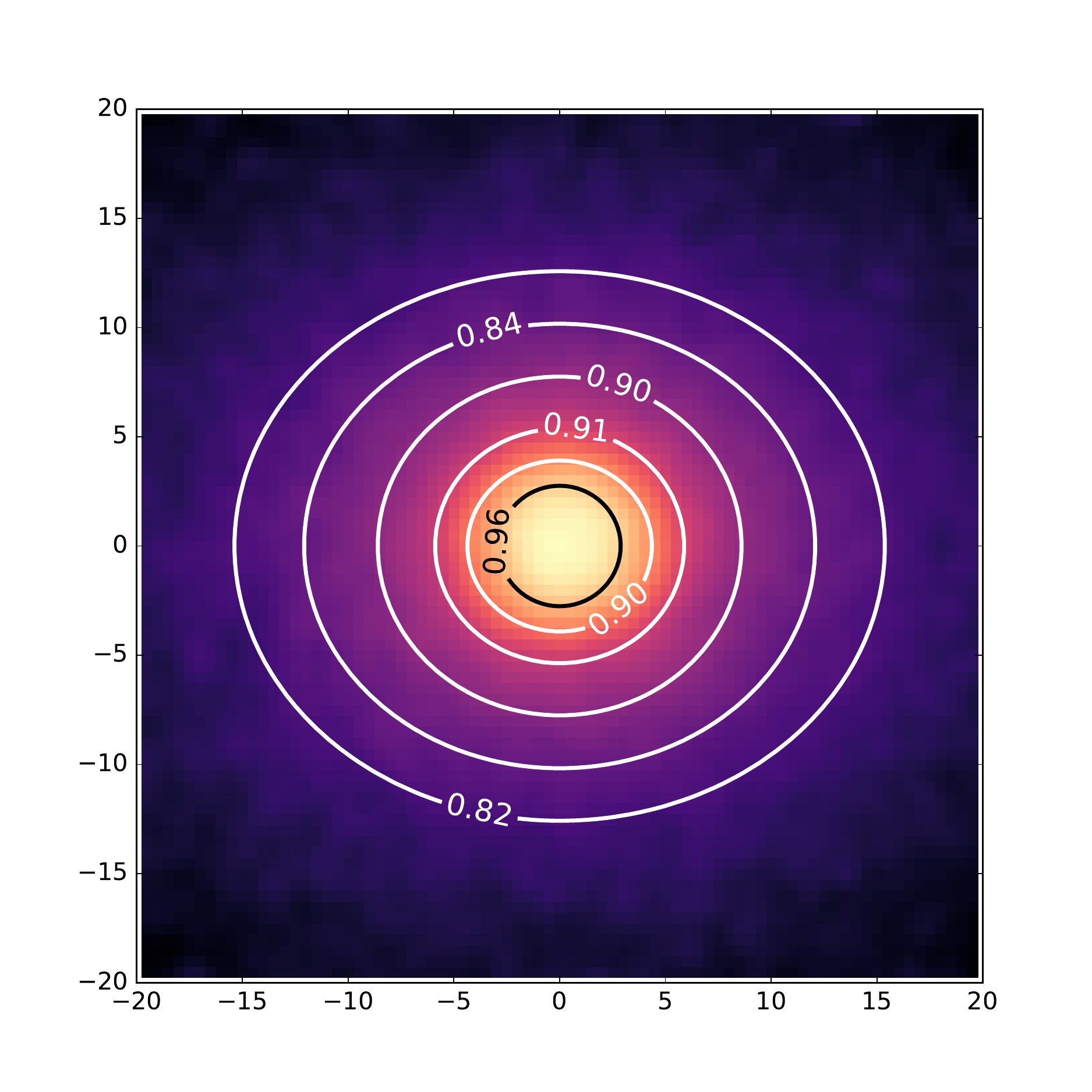}\\
\caption{Projected annihilation surface brightness map of the galactic centre region (a $40^{\circ}\times40^{\circ}$ patch viewed from a solar-like position) in one of our level 3 full physics simulations. The projected galactic disk is horizontal. The colour scale is logarithmic with arbitrary zeropoint. Ellipses are fit to  logarithmically-spaced isophotes (white and black contours). The vertical-to-horizontal axis ratio, $q$,  is indicated for each fitted ellipse.}
\label{maps}
\end{figure}

\begin{figure}
\includegraphics[width=\columnwidth,trim={0.2cm 0.8cm 0.2cm 0.8cm}, clip]{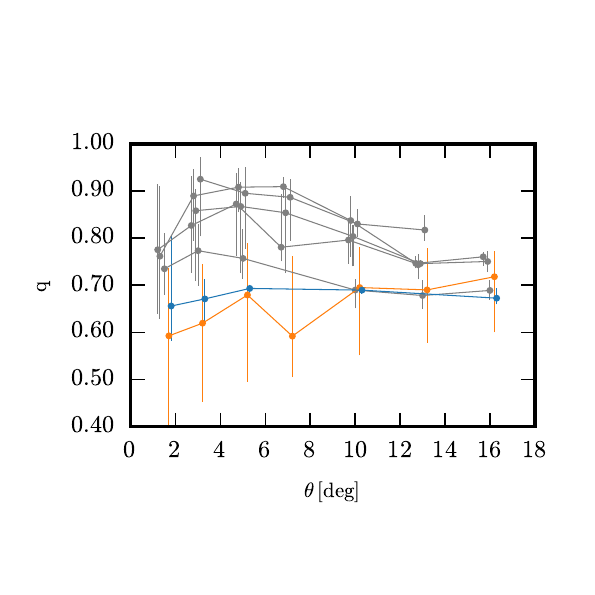}\\
\caption{The axial ratio, $q$, as a function of angular distance from the galactic centre, for all 18 solar-like positions in each simulation. Circles show the median $q$ values in each angular bin with vertical bars to indicate the full range of $q$ across all solar-like positions. The $x$-axis positions of points and their associated bars are offset by small amounts for clarity. Colours are as in Fig.~\ref{ASB}. Most simulations are consistent with $q\gtrsim 0.8$ for $\theta \lesssim 12^{\circ}$.}
\label{ellvar}
\end{figure}

\begin{figure}
\includegraphics[width=\columnwidth,trim={11cm 0 5cm 0}, clip]{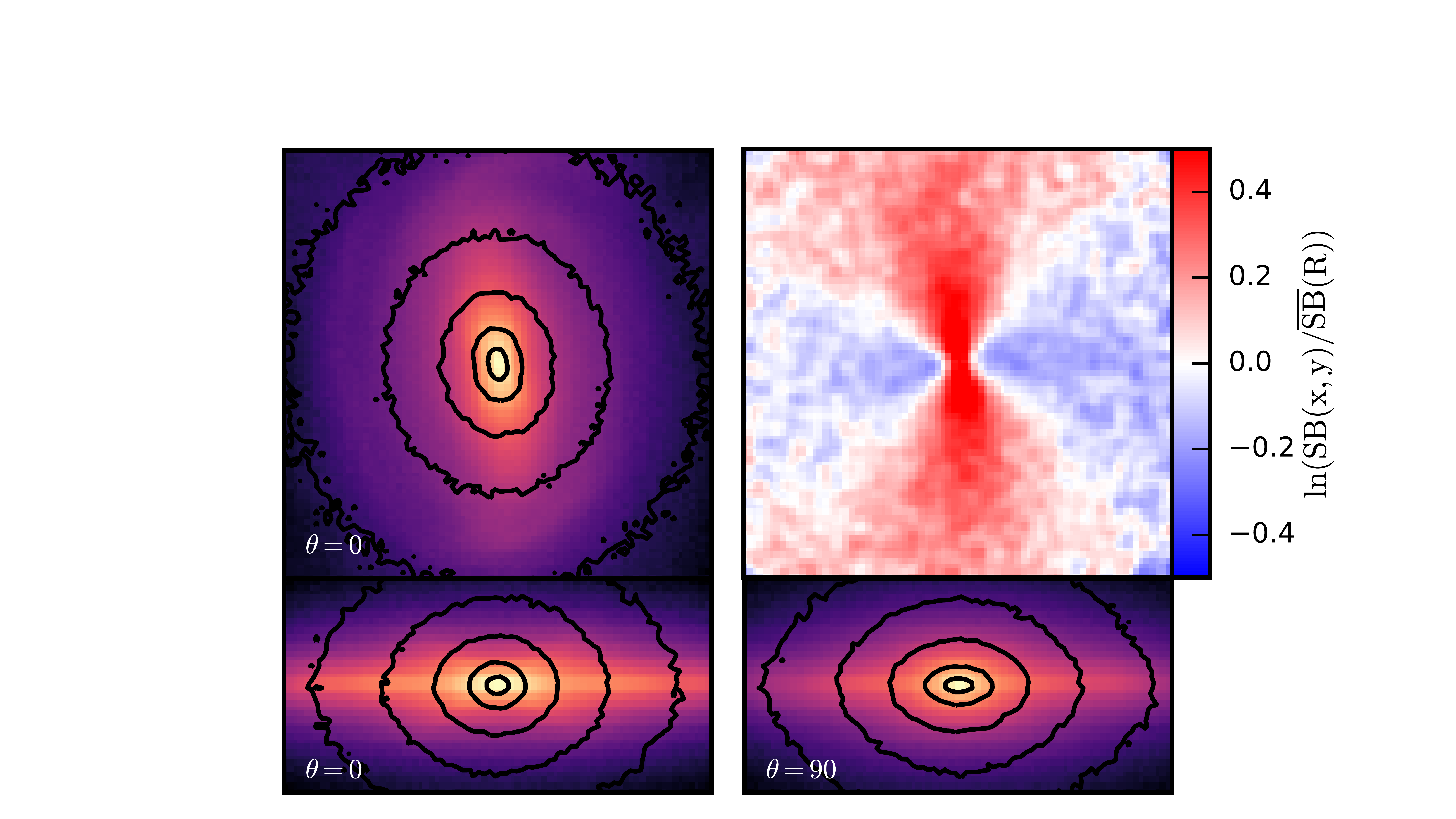}\\
\caption{Top left: Contours of annihilation surface brightness (ASB) as seen by a distant observer towards the galactic pole are overlaid on an image of the projected stellar mass distribution for a $20{\rm kpc}\times 20{\rm kpc}$ region of the simulation indicated in orange in Figures \ref{ASB} and \ref{ellvar}. Top right: The relative azimuthal variation in ASB at constant galactocentric radius in the same projected image. Bottom panels: ASB contours and the projected stellar mass distribution in the same format and to the same scale as at top left but now for edge-on projections along (left) and transverse to (right) the stellar bar.}
\label{darkbar}
\end{figure}

The morphologyy of the GCE is an additional diagnostic that can help interpret the source of the signal. \citet{DiMauro21} found the GCE to be approximately circularly symmetric about the Galactic Centre, which would be
consistent with DM annihilation from a near-spherical halo. He found this morphology to be clearly preferred over one reflecting production from a population of stars or gas clouds associated with the disc, which would produce a signal strongly elongated along the Galactic Plane. 

We quantify the morphology of DM annihilation maps made from our simulations by fitting ellipses to the isophotes of the surface brightness in each of the 18 projections of each simulation. The ellipses are centred on the galactic centre with axes aligned with the disc midplane and the rotation axis. As is conventional, we define the flattening $q$ as the ratio of the axis aligned with the rotation axis to that aligned with the disc. Fig.~\ref{maps} shows an example of the resulting ASB maps for a $40^{\circ}\times40^{\circ}$ patch surrounding the galactic centre. The fitted ellipses for various ASB values are shown as white (or black) contours labelled by their $q$ values. The rms percentage error in surface brightness around the fitted ellipses in this and almost all similar plots for this and other simulations is between $5\%$ and $10 \%$. 

In Fig.\ref{ellvar} we show the variation of $q$ with radius for different vantage points in all of our simulations. The colours correspond to those in Fig.\ref{ASB}. We exclude vantage points for which the surface brightness map contains clear ``blob-like'' features because this can distort some of the fitted ellipses quite far from the smooth surface brightness isophotes. Fewer than a third of the vantage points are typically excluded in each simulation. For the real GCE, \citet{DiMauro21} finds best values for $q$ in the range 0.8 to 1.2 depending on the specific model adopted for the foregrounds (note that his convention for the axial ratio is the reciprocal of ours). This is consistent with the results found at most radii in most of our simulations where the overall median value of $q$ is about 0.8.
It is noticeable that the most flattened case, and also the one with the largest variation in $q$ with vantage point is the one shown in orange, the level 3 counterpart of our single high-resolution simulation. 

In Fig.~\ref{darkbar}, we demonstrate that this variability is caused by a strong stellar bar in this simulation, which can be seen very clearly in Fig.1 of \citet{vdV2021}. The top left panel of our figure overlays contours of the ASB on an image of the projected stellar mass distribution in the inner 10 kpc of the galaxy. Clearly, the DM distribution is not axisymmetric but rather is elongated along the stellar bar. As can be seen in the top right panel, this ellipsoidal distortion is strongest near the centre, reaching amplitudes of almost a factor of two in ASB,  but is present over a wide range of radii. The lower panels of Fig.~\ref{darkbar} show edge-on views projected along (left) and perpendicular to (right) the bar. These highlight the variation in isophote flattening with viewing angle responsible for the substantial scatter around the orange points in Fig. \ref{ellvar}. In these views it is noticeable that the isophotes are lens-shaped rather than exactly elliptical. The bar is weaker in the high-resolution version of this simulation, and, although still strongly flattened, it shows much less variation in $q$ with viewing angle.

\section{Conclusions}
\label{conclusions}

The Fermi-LAT data agree well with the simulated  shape both of the radial ASB profile and of the ASB isophotes over quite a large angular range, 0-15$^{\circ}$. This suggests that the GCE may be dominated by emission from DM annihilation. If, however, it is due to a different source, for example, to a bulge population of millisecond pulsars, then our results imply that the observations can be used to put a strong upper limit on the ``particle physics factor'' $\mathcal{C}$ in our equation \ref{lumden} for the luminosity density, translating into a strong upper limit on the $\gamma$-ray annihilation cross-section of DM. In his Fig.10, \citet{DiMauro21} gives the surface brightness of the GCE in units of MeV/cm$^2$/s/sr. Using these same energy units our results give the upper limit 
\begin{equation}
f\frac{\langle \sigma v\rangle c^2}{m_{\chi}} = \mathcal{C} < 1.17\times 10^{16}~~ {\rm MeV~cm^3/s/g^2},
\end{equation}
where $f$ is the fraction of annihilation energy of WIMPs of mass $m_{\chi}$ and ``thermally'' averaged pair-wise annihilation cross-section $\langle \sigma v\rangle$ which appears as $\gamma$-rays in the Fermi-LAT 1--10 GeV band.
As detailed in our earlier paper \citep{GW21}, such an upper limit on the annihilation flux from the inner Galaxy implies an upper limit on the expected flux from any substructure (with or without visible stars) in the Galactic halo which is more stringent than assumed in previous work because of the substantially enhanced contrast between the GCE and satellite emission resulting from baryonic effects..

Our magnetohydrodynamical ``full physics'' simulations, carried out within the standard $\Lambda$CDM paradigm, produce galaxies in quantitative agreement with many aspects of the observed structure of our Milky Way. The fact that they simultaneously and without further adjustment also reproduce the morphology and the shape of the Fermi GCE supports the hypothesis that this extended component of $\gamma$-ray emission is indeed a product of dark matter annihilation. The most significant outstanding challenge to this interpretation remains the relatively strong smaller scale fluctuations in the observed GCE surface brightness \citep[see Fig.5 of][]{DiMauro21}. These may favour production by a population of point-sources such as millisecond pulsars, or they may reflect residual small-scale errors in the foreground templates \citep[see][for some discussion of the issues]{Slatyer2021}.

\section*{Acknowledgements}
RG acknowledges financial support from the Spanish Ministry of Science and Innovation (MICINN) through the Spanish State Research Agency, under the Severo Ochoa Program 2020-2023 (CEX2019-000920-S).

\section*{Data Availability}
The data underlying this article will be shared on reasonable request to the corresponding author.

\bibliographystyle{mnras}
\bibliography{pap1.bib}

\newcommand{\noop}[1]{}
\begin{thebibliography}{}
\makeatletter
\relax
\def\mn@urlcharsother{\let\do\@makeother \do\$\do\&\do\#\do\^\do\_\do\%\do\~}
\def\mn@doi{\begingroup\mn@urlcharsother \@ifnextchar [ {\mn@doi@}
  {\mn@doi@[]}}
\def\mn@doi@[#1]#2{\def\@tempa{#1}\ifx\@tempa\@empty \href
  {http://dx.doi.org/#2} {doi:#2}\else \href {http://dx.doi.org/#2} {#1}\fi
  \endgroup}
\def\mn@eprint#1#2{\mn@eprint@#1:#2::\@nil}
\def\mn@eprint@arXiv#1{\href {http://arxiv.org/abs/#1} {{\tt arXiv:#1}}}
\def\mn@eprint@dblp#1{\href {http://dblp.uni-trier.de/rec/bibtex/#1.xml}
  {dblp:#1}}
\def\mn@eprint@#1:#2:#3:#4\@nil{\def\@tempa {#1}\def\@tempb {#2}\def\@tempc
  {#3}\ifx \@tempc \@empty \let \@tempc \@tempb \let \@tempb \@tempa \fi \ifx
  \@tempb \@empty \def\@tempb {arXiv}\fi \@ifundefined
  {mn@eprint@\@tempb}{\@tempb:\@tempc}{\expandafter \expandafter \csname
  mn@eprint@\@tempb\endcsname \expandafter{\@tempc}}}

\bibitem[\protect\citeauthoryear{{Ackermann} et~al.,}{{Ackermann}
  et~al.}{2012}]{Ackermann+12}
{Ackermann} M.,  et~al., 2012, \mn@doi [\apj] {10.1088/0004-637X/761/2/91},
  \href {https://ui.adsabs.harvard.edu/abs/2012ApJ...761...91A} {761, 91}

\bibitem[\protect\citeauthoryear{{Albert} et~al.,}{{Albert}
  et~al.}{2017}]{Albert+17}
{Albert} A.,  et~al., 2017, \mn@doi [\apj] {10.3847/1538-4357/834/2/110}, \href
  {https://ui.adsabs.harvard.edu/abs/2017ApJ...834..110A} {834, 110}

\bibitem[\protect\citeauthoryear{{Arcadi}, {Dutra}, {Ghosh}, {Lindner},
  {Mambrini}, {Pierre}, {Profumo}  \& {Queiroz}}{{Arcadi}
  et~al.}{2018}]{Arcadi+18}
{Arcadi} G.,  {Dutra} M.,  {Ghosh} P.,  {Lindner} M.,  {Mambrini} Y.,  {Pierre}
  M.,  {Profumo} S.,   {Queiroz} F.~S.,  2018, \mn@doi [European Physical
  Journal C] {10.1140/epjc/s10052-018-5662-y}, \href
  {https://ui.adsabs.harvard.edu/abs/2018EPJC...78..203A} {78, 203}

\bibitem[\protect\citeauthoryear{{Atwood} et~al.,}{{Atwood}
  et~al.}{2009}]{Fermi09}
{Atwood} W.~B.,  et~al., 2009, \mn@doi [\apj] {10.1088/0004-637X/697/2/1071},
  \href {https://ui.adsabs.harvard.edu/abs/2009ApJ...697.1071A} {697, 1071}

\bibitem[\protect\citeauthoryear{{Berezinsky}, {Bottino}  \&
  {Mignola}}{{Berezinsky} et~al.}{1994}]{BBM94}
{Berezinsky} V.,  {Bottino} A.,   {Mignola} G.,  1994, \mn@doi [Physics Letters
  B] {10.1016/0370-2693(94)90083-3}, \href
  {https://ui.adsabs.harvard.edu/abs/1994PhLB..325..136B} {325, 136}

\bibitem[\protect\citeauthoryear{{Bergstr{\"o}m}, {Ullio}  \&
  {Buckley}}{{Bergstr{\"o}m} et~al.}{1998}]{BUB98}
{Bergstr{\"o}m} L.,  {Ullio} P.,   {Buckley} J.~H.,  1998, \mn@doi
  [Astroparticle Physics] {10.1016/S0927-6505(98)00015-2}, \href
  {https://ui.adsabs.harvard.edu/abs/1998APh.....9..137B} {9, 137}

\bibitem[\protect\citeauthoryear{{Bertone}, {Hooper}  \& {Silk}}{{Bertone}
  et~al.}{2005}]{BHS05}
{Bertone} G.,  {Hooper} D.,   {Silk} J.,  2005, \mn@doi [\physrep]
  {10.1016/j.physrep.2004.08.031}, \href
  {https://ui.adsabs.harvard.edu/abs/2005PhR...405..279B} {405, 279}

\bibitem[\protect\citeauthoryear{{Cautun} et~al.,}{{Cautun}
  et~al.}{2020}]{CBD20}
{Cautun} M.,  et~al., 2020, \mn@doi [\mnras] {10.1093/mnras/staa1017}, \href
  {https://ui.adsabs.harvard.edu/abs/2020MNRAS.494.4291C} {494, 4291}

\bibitem[\protect\citeauthoryear{{Chang}, {Lisanti}  \&
  {Mishra-Sharma}}{{Chang} et~al.}{2018}]{CLM18}
{Chang} L.~J.,  {Lisanti} M.,   {Mishra-Sharma} S.,  2018, \mn@doi [\prd]
  {10.1103/PhysRevD.98.123004}, \href
  {https://ui.adsabs.harvard.edu/abs/2018PhRvD..98l3004C} {98, 123004}

\bibitem[\protect\citeauthoryear{{Di Mauro}}{{Di Mauro}}{2021}]{DiMauro21}
{Di Mauro} M.,  2021, \mn@doi [\prd] {10.1103/PhysRevD.103.063029}, \href
  {https://ui.adsabs.harvard.edu/abs/2021PhRvD.103f3029D} {103, 063029}

\bibitem[\protect\citeauthoryear{{Fragkoudi} et~al.,}{{Fragkoudi}
  et~al.}{2020}]{Fragkoudi+Grand+Pakmor+19}
{Fragkoudi} F.,  et~al., 2020, \mn@doi [\mnras] {10.1093/mnras/staa1104}, \href
  {https://ui.adsabs.harvard.edu/abs/2020MNRAS.494.5936F} {494, 5936}

\bibitem[\protect\citeauthoryear{{Gargiulo} et~al.,}{{Gargiulo}
  et~al.}{2019}]{GMG19}
{Gargiulo} I.~D.,  et~al., 2019, \mn@doi [\mnras] {10.1093/mnras/stz2536},
  \href {https://ui.adsabs.harvard.edu/abs/2019MNRAS.489.5742G} {489, 5742}

\bibitem[\protect\citeauthoryear{{Gaskins}}{{Gaskins}}{2016}]{G16}
{Gaskins} J.~M.,  2016, \mn@doi [Contemporary Physics]
  {10.1080/00107514.2016.1175160}, \href
  {https://ui.adsabs.harvard.edu/abs/2016ConPh..57..496G} {57, 496}

\bibitem[\protect\citeauthoryear{{Grand} \& {White}}{{Grand} \&
  {White}}{2021}]{GW21}
{Grand} R. J.~J.,  {White} S. D.~M.,  2021, \mn@doi [\mnras]
  {10.1093/mnras/staa3993}, \href
  {https://ui.adsabs.harvard.edu/abs/2021MNRAS.501.3558G} {501, 3558}

\bibitem[\protect\citeauthoryear{{Grand} et~al.,}{{Grand} et~al.}{2017}]{GGM17}
{Grand} R.~J.~J.,  et~al., 2017, \mn@doi [\mnras] {10.1093/mnras/stx071}, \href
  {http://adsabs.harvard.edu/abs/2017MNRAS.467..179G} {467, 179}

\bibitem[\protect\citeauthoryear{{Grand} et~al.,}{{Grand}
  et~al.}{2018a}]{GBG18}
{Grand} R.~J.~J.,  et~al., 2018a, \mn@doi [\mnras] {10.1093/mnras/stx3025},
  \href {http://adsabs.harvard.edu/abs/2018MNRAS.474.3629G} {474, 3629}

\bibitem[\protect\citeauthoryear{{Grand}, {Helly}, {Fattahi}  et~al.}{{Grand}
  et~al.}{2018b}]{GHF18}
{Grand} R.~J.~J.,  {Helly} J.,  {Fattahi} A.,   et~al., 2018b, \mn@doi [\mnras]
  {10.1093/mnras/sty2403}, \href
  {http://adsabs.harvard.edu/abs/2018MNRAS.481.1726G} {481, 1726}

\bibitem[\protect\citeauthoryear{{Grand} et~al.,}{{Grand} et~al.}{2020}]{GKB20}
{Grand} R. J.~J.,  et~al., 2020, \mn@doi [\mnras] {10.1093/mnras/staa2057},
  \href {https://ui.adsabs.harvard.edu/abs/2020MNRAS.497.1603G} {497, 1603}

\bibitem[\protect\citeauthoryear{{Grand} et~al.,}{{Grand} et~al.}{2021}]{GMP21}
{Grand} R. J.~J.,  et~al., 2021, \mn@doi [\mnras] {10.1093/mnras/stab2492},
  \href {https://ui.adsabs.harvard.edu/abs/2021MNRAS.507.4953G} {507, 4953}

\bibitem[\protect\citeauthoryear{{Hooper} \& {Goodenough}}{{Hooper} \&
  {Goodenough}}{2011}]{Hooper2011}
{Hooper} D.,  {Goodenough} L.,  2011, \mn@doi [Physics Letters B]
  {10.1016/j.physletb.2011.02.029}, \href
  {https://ui.adsabs.harvard.edu/abs/2011PhLB..697..412H} {697, 412}

\bibitem[\protect\citeauthoryear{{Lee}, {Lisanti}, {Safdi}, {Slatyer}  \&
  {Xue}}{{Lee} et~al.}{2016}]{Lee2016}
{Lee} S.~K.,  {Lisanti} M.,  {Safdi} B.~R.,  {Slatyer} T.~R.,   {Xue} W.,
  2016, \mn@doi [\prl] {10.1103/PhysRevLett.116.051103}, \href
  {https://ui.adsabs.harvard.edu/abs/2016PhRvL.116e1103L} {116, 051103}

\bibitem[\protect\citeauthoryear{{Lovell} et~al.,}{{Lovell}
  et~al.}{2018}]{LPG18}
{Lovell} M.~R.,  et~al., 2018, \mn@doi [\mnras] {10.1093/mnras/sty2339}, \href
  {https://ui.adsabs.harvard.edu/abs/2018MNRAS.481.1950L} {481, 1950}

\bibitem[\protect\citeauthoryear{{Marinacci}, {Grand}, {Pakmor}, {Springel},
  {G{\'o}mez}, {Frenk}  \& {White}}{{Marinacci} et~al.}{2017}]{MGP16}
{Marinacci} F.,  {Grand} R.~J.~J.,  {Pakmor} R.,  {Springel} V.,  {G{\'o}mez}
  F.~A.,  {Frenk} C.~S.,   {White} S.~D.~M.,  2017, \mn@doi [\mnras]
  {10.1093/mnras/stw3366}, \href
  {http://adsabs.harvard.edu/abs/2017MNRAS.466.3859M} {466, 3859}

\bibitem[\protect\citeauthoryear{{Planck Collaboration}}{{Planck
  Collaboration}}{2014}]{PC13}
{Planck Collaboration} 2014, \mn@doi [\aap] {10.1051/0004-6361/201321591},
  \href {http://adsabs.harvard.edu/abs/2014A%26A...571A..16P} {571, A16}

\bibitem[\protect\citeauthoryear{{Schaller} et~al.,}{{Schaller}
  et~al.}{2015}]{SFB15}
{Schaller} M.,  et~al., 2015, \mn@doi [\mnras] {10.1093/mnras/stv1067}, \href
  {http://adsabs.harvard.edu/abs/2015MNRAS.451.1247S} {451, 1247}

\bibitem[\protect\citeauthoryear{{Slatyer}}{{Slatyer}}{2021}]{Slatyer2021}
{Slatyer} T.~R.,  2021, arXiv e-prints, \href
  {https://ui.adsabs.harvard.edu/abs/2021arXiv210902696S} {p. arXiv:2109.02696}

\bibitem[\protect\citeauthoryear{{Springel}}{{Springel}}{2010}]{Sp10}
{Springel} V.,  2010, \mn@doi [\mnras] {10.1111/j.1365-2966.2009.15715.x},
  \href {http://adsabs.harvard.edu/abs/2010MNRAS.401..791S} {401, 791}

\bibitem[\protect\citeauthoryear{{Springel} et~al.,}{{Springel}
  et~al.}{2008}]{SWF08}
{Springel} V.,  et~al., 2008, \mn@doi [\nat] {10.1038/nature07411}, \href
  {https://ui.adsabs.harvard.edu/abs/2008Natur.456...73S} {456, 73}

\bibitem[\protect\citeauthoryear{{Strigari}, {Koushiappas}, {Bullock}  \&
  {Kaplinghat}}{{Strigari} et~al.}{2007}]{Strigari2007}
{Strigari} L.~E.,  {Koushiappas} S.~M.,  {Bullock} J.~S.,   {Kaplinghat} M.,
  2007, \mn@doi [\prd] {10.1103/PhysRevD.75.083526}, \href
  {https://ui.adsabs.harvard.edu/abs/2007PhRvD..75h3526S} {75, 083526}

\bibitem[\protect\citeauthoryear{{Wang}, {Bose}, {Frenk}, {Gao}, {Jenkins},
  {Springel}  \& {White}}{{Wang} et~al.}{2020}]{WBF20}
{Wang} J.,  {Bose} S.,  {Frenk} C.~S.,  {Gao} L.,  {Jenkins} A.,  {Springel}
  V.,   {White} S.~D.~M.,  2020, \mn@doi [\nat] {10.1038/s41586-020-2642-9},
  \href {https://ui.adsabs.harvard.edu/abs/2020Natur.585...39W} {585, 39}

\bibitem[\protect\citeauthoryear{{van de Voort}, {Pakmor}, {Bieri}  \&
  {Grand}}{{van de Voort} et~al.}{2021}]{vdV2021}
{van de Voort} F.,  {Pakmor} R.,  {Bieri} R.,   {Grand} R. J.~J.,  2021, arXiv
  e-prints, \href {https://ui.adsabs.harvard.edu/abs/2021arXiv211011963V} {p.
  arXiv:2110.11963}

\makeatother
\end{thebibliography}

\bsp	
\label{lastpage}
\end{document}